\documentclass[12pt,a4paper]{article}
\usepackage[latin1]{inputenc}
\usepackage{amsmath}
\usepackage{amssymb}
\usepackage{amsthm}
\usepackage{geometry}
\usepackage{booktabs}
\usepackage{pgfplots}
\pgfplotsset{compat=1.16}
\usepgfplotslibrary{fillbetween}
\usetikzlibrary{arrows.meta}
\usepackage{natbib}
\usepackage{placeins}

\newtheorem{prop}{Proposition}

\theoremstyle{remark}
\newtheorem{rem}{Remark}

\begin{document}

\begin{center}\huge{AI-Augmented Research:\\ Free to Write, Costly to Read}\footnote{I thank Sabina Karabaeva for helpful discussions.}
\end{center}

\begin{center} \textit{Wolfgang Kuhle}\\ \textit{Corvinus University of Budapest, Hungary, E-mail wkuhle@gmx.de\\ MEA, Max Planck Institute for Social Law and Social Policy, Munich, Germany}\end{center}

\noindent\emph{\textbf{Abstract:} We study a model in which artificial intelligence lowers the cost of writing faster than the cost of reading. Authors differ in reputation, and time-constrained readers, such as referees, choose between ignoring, verifying, and trusting papers. As writing costs fall, the band of verified papers shrinks: outsiders are no longer read, and established authors are no longer checked. In the ``costless writing limit,'' the share of verified papers vanishes, and a paper's impact depends on its author's reputation alone. Readers can escape the flood by asking AI to summarize the literature. These summaries aggregate papers reliably only if their errors are sufficiently independent. Once the share of papers written with the same AI exceeds a threshold, summaries inherit the AI's errors, and AI review filters out the original research that could correct them.}

\section{Introduction}\label{Section Introduction}

Writing a page of research used to be expensive. Modern large language models have made it nearly free. At the same time, understanding and verifying a proof, a data set, or a calibration still takes time. This paper asks what happens to the market for research when the cost of writing falls faster than the cost of verification.

Our model is populated by authors and readers. Authors differ in reputation, the probability that their research is correct, and readers allocate limited attention. For each paper, readers can ignore it, verify it at a cost, or trust it without checking. Authors benefit from producing papers and from having their work used. As writing becomes cheaper, they produce more papers, but the attention available for verification remains fixed.

Building on this model, Section \ref{Section Equilibrium} shows that, in equilibrium, readers verify papers by authors of intermediate reputation, ignore those with low reputation, and trust those with high reputation. As writing becomes cheaper, more papers compete for the same attention, making verification increasingly costly. The range of authors whose papers are verified shrinks: outsiders are no longer read, and established authors are no longer checked. In the costless writing limit, only a vanishing share of papers is verified, and a paper is used if and only if its author's reputation exceeds a fixed threshold.

Section \ref{Section Summaries} studies whether AI summaries can overcome the scarcity of attention. A summary reporting the majority view of a literature works like the jury of \citet{Condorcet1785}: it recovers the correct answer when sufficiently many papers reach independent, better-than-random judgments. Papers written with the same AI, however, share its errors. Summaries thus overcome the attention bottleneck but create a new problem of correlated errors. 

In an alternative interpretation, our model studies journals' editorial policies: ignoring corresponds to a desk rejection, verification to refereeing, and trusting without verification to acceptance on reputation. Taking this view, Section \ref{Section Review} examines what happens when the AI also acts as a referee. An AI that favors papers agreeing with its own answer filters out dissenting research. When the AI is wrong, these are precisely the papers that could correct its error. AI review, therefore, lowers the threshold at which a literature's consensus is captured. If the AI is sufficiently resistant to disagreement, its answers dominate the literature even when no paper is written with it.

Section \ref{Section Policy} shows that the simplest remedy is an unrefereed preprint server from which summaries draw. It restores a correct consensus as long as AI-written papers stay below a certain threshold. %It cannot, however, overcome the underlying dependence of AI-written papers: once their share exceeds this threshold, no acceptance rule can guarantee a correct consensus.

\textit{Related literature.} \citet{Simon1971} observed that a wealth of information creates a poverty of attention. Related, \citet{DasguptaDavid1994} argue that reputation drives the attention that different research results get. The growth in the number of papers that motivates our model is documented by \citet{Hanson2024}. \citet{CardDellaVigna2013} document the rising burden of refereeing in economics, and \citet{BergstromGross2025} show that review noise that rises with submission volume can push peer review into collapse.

\citet{Liang2025} and \citet{Kobak2025} measure the rising share of scientific text written with large language models. \citet{Russo2024} find that AI-assisted reviews raise paper scores and acceptance rates. Our result on filtered originality relates to the evidence of \citet{Wang2017} that novel research is penalized. Our results on AI summaries and AI review relate to the evidence of \citet{DoshiHauser2024} that generative AI reduces the collective diversity of novel content, and of \citet{Qian2026} that grant proposals written with heavy use of large language models are less distinct from previously funded work. \citet{Ladha1992} shows that correlated votes undermine the jury theorem, and \citet{Zollman2007} shows that scientific communities can be more reliable when their members share less.

Relative to this literature, we derive how cheap writing leads to two failures. When writing becomes cheaper than verification, scarce attention is rationed by reputation rather than spread evenly as in \citet{BergstromGross2025}, and the ratio of the two costs decides whether AI helps unknown/low reputation authors to be read. When readers and journals turn to AI, errors that are correlated through a common model capture the consensus above an explicit share of AI-written papers, and AI review that screens on agreement lowers this share further because it removes the evidence that would correct the AI.

\section{Model}\label{Section Model}

\textit{Authors.} A continuum of authors of unit mass differs in reputation,\footnote{Reputation may, in particular, be derived from the work that the author has published prior to the AI research flood.} $\rho\in[0,1]$, which is the probability that a paper by the author is correct. Reputations have a continuous density $f$ with $f>0$ on $(0,1)$ and distribution function $F$. An author who writes $n$ papers bears the cost $c_wn^2/2$, where $c_w>0$ measures the cost of writing; free writing refers to the limit $c_w\to0$. Each paper yields a private benefit $b_0>0$, for example from publication counts in hiring,\footnote{The private benefit ensures that ignored authors also write. With $b_0=0$, ignored authors write nothing, but the results below on the verified and trusted bands are unchanged.} and an additional benefit $b>0$ if readers use it. An author with use probability $u$ per paper therefore writes
\begin{equation}\label{Eq Papers}
n(\rho)=\frac{b_0+b\,u(\rho)}{c_w}
\end{equation}
papers. Authors are small and take readers' behavior as given.

\textit{Readers.} Using a correct result yields value $V>0$, and using an incorrect one causes damage $D>0$. For each paper, readers can ignore it, with payoff $0$; trust it without verification, with payoff $\rho V-(1-\rho)D$; or verify it and use it only if it is correct, with payoff $\rho V-c_v$.\footnote{\label{Fn Imperfect}Perfect verification is a benchmark; \citet{Bornmann2010} find that referees agree only modestly with each other. If verification detects an error only with probability $\eta<1$, its payoff becomes $\rho V-(1-\rho)(1-\eta)D-c_v$. This payoff still falls in $c_v$, while the payoffs of ignoring and trusting do not, so that the verified band still shrinks from both sides as attention becomes scarce.} Verification stands for checking proofs, reproducing code and data analyses, or replicating results. It requires $\kappa>0$ units of attention, and readers have $A>0$ units in total, which we interpret as the attention of a field.\footnote{Readers and authors may be the same researchers. If time spent writing reduces the attention available for verification, cheaper writing also lowers $A$, which reinforces the results below.} The cost of verification is $c_v=\lambda\kappa$, where $\lambda\ge0$ is the shadow price of attention. Ignoring and trusting take no attention: $\kappa$ is the attention that checking a paper takes beyond using it, and we normalize the attention of using a paper to zero. Use probabilities are $u=0$ for ignored papers, $u=\rho$ for verified papers, and $u=1$ for trusted papers.

\textit{Equilibrium.} An equilibrium consists of a shadow price $\lambda$ and a paper supply $n(\rho)$ such that authors write (\ref{Eq Papers}) given readers' choices, readers choose optimally given $c_v=\lambda\kappa$, the attention used for verification does not exceed $A$, and $\lambda=0$ if it falls short of $A$.

\section{Equilibrium}\label{Section Equilibrium}

All proofs are in Appendix \ref{Appendix Proofs}. Let $\bar c:=VD/(V+D)$ and $\bar\rho:=D/(V+D)$.

\begin{prop}[Reading Rule]\label{Prop Reading}
If $c_v<\bar c$, readers ignore papers with $\rho<\rho_L:=c_v/V$, verify papers with $\rho_L\le\rho<\rho_H:=1-c_v/D$, and trust papers with $\rho\ge\rho_H$. If $c_v\ge\bar c$, readers verify no paper, ignore papers with $\rho<\bar\rho$, and trust papers with $\rho\ge\bar\rho$.
\end{prop}

Proposition \ref{Prop Reading} indicates that verification pays only for authors of intermediate reputation. A reader verifies rather than ignores if the expected value of a correct result exceeds the cost of checking, $\rho V\ge c_v$, and verifies rather than trusts if the expected damage from an incorrect result exceeds it, $(1-\rho)D\ge c_v$. A higher cost of verification shrinks the verified band from both sides, and at $c_v=\bar c$ the band collapses to the single reputation $\bar\rho$ at which trusting and ignoring yield the same payoff.

Define the attention demand
$$\Phi(c_v):=\int_{c_v/V}^{1-c_v/D}(b_0+b\rho)f(\rho)\,d\rho,$$
which is continuous and strictly decreasing on $[0,\bar c]$, with $\Phi(\bar c)=0$. By (\ref{Eq Papers}), the attention used for verification is $\kappa\Phi(c_v)/c_w$.

\begin{prop}[Equilibrium]\label{Prop Equilibrium}
There is a unique equilibrium. If $c_w\ge\kappa\Phi(0)/A$, attention is slack, $c_v=0$, and every paper is verified. If $c_w<\kappa\Phi(0)/A$, the cost of verification $c_v\in(0,\bar c)$ solves
\begin{equation}\label{Eq Equilibrium}
\Phi(c_v)=\frac{A\,c_w}{\kappa},
\end{equation}
and $c_v$ rises as $c_w$ falls. As $c_w\to0$, $c_v\to\bar c$, both cutoffs $\rho_L$ and $\rho_H$ converge to $\bar\rho$, the number of papers $N$ grows like $1/c_w$, while the number of verified papers stays at $A/\kappa$, so that the share of papers that are verified converges to zero.
\end{prop}

In the limit, the decision to use a paper rests on its author's reputation alone: papers by authors above $\bar\rho$ are believed, and all others are ignored.

\begin{rem}[Peer Review]\label{Rem PeerReview}
The reading rule also describes a journal. Ignoring means desk rejection, verification means refereeing, and trust means acceptance based on reputation or lighter review for established authors. Likewise, $A$ would represent the size of the referee pool and $\kappa$ the time a report takes. As submissions rise, refereeing is confined to authors of intermediate reputation: outsiders are desk-rejected, and insiders are waved through.\footnote{This observation matches the experiment of \citet{Huber2022}: when a paper listed only its prominent co-author, a Nobel laureate, $23\%$ of referees recommended rejection, compared with $48\%$ when it was anonymous and $65\%$ when it listed only a junior co-author.}
\end{rem}

\begin{prop}[The Race between Writing and Reading]\label{Prop Race}
The equilibrium cost of verification, the cutoffs $\rho_L$ and $\rho_H$, and the set of verified reputations depend on $(c_w,\kappa,A)$ only through $Ac_w/\kappa$. If attention binds and technology changes $c_w$ and $\kappa$, the cutoff $\rho_L$ for being read falls, and the set of verified reputations widens if and only if $\kappa$ falls proportionally faster than $c_w$, holding $A$ fixed.
\end{prop}

Proposition \ref{Prop Race} states the condition under which AI helps outsiders, in the sense that authors of lower reputation are read.\footnote{In terms of utility, an author with reputation $\rho$ obtains $(b_0+b\,u(\rho))^2/(2c_w)$. Cheaper writing raises this payoff for every author, whether read or not, through the private benefit $b_0$. What technology changes for outsiders is whether they are read at all: an author whom the fall in $\rho_L$ moves from the ignored into the verified band gains the readership benefit $b\rho$ on each paper.}

\subsection{Diagrammatic Interpretation}\label{Section Diagrams}

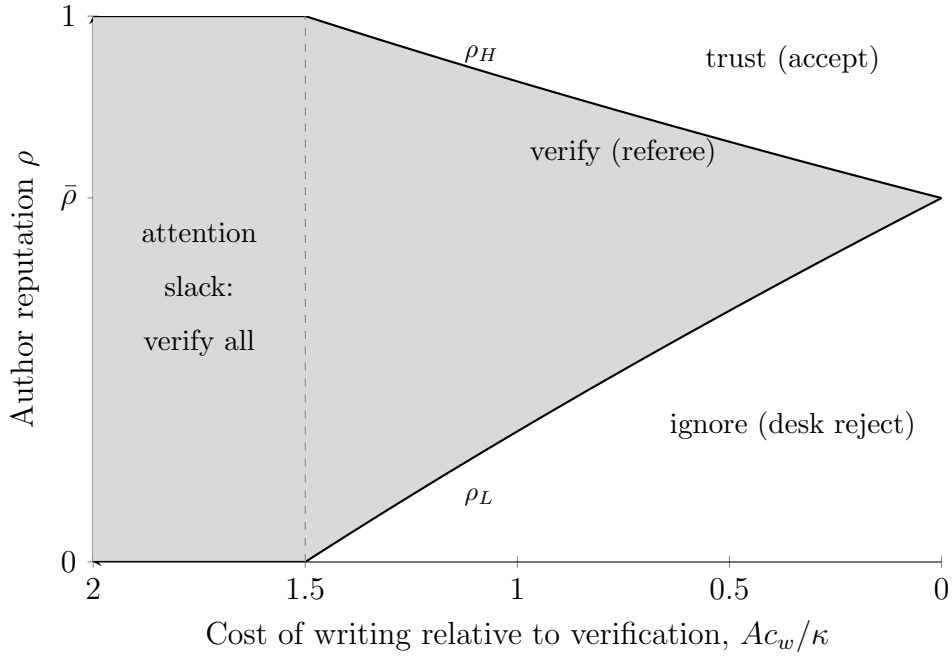
\begin{figure}[h!]
\centering
\begin{tikzpicture}
\begin{axis}[width=0.8\textwidth,height=0.55\textwidth,
  x dir=reverse,xmin=0,xmax=2,ymin=0,ymax=1,axis lines=left,
  xlabel={Cost of writing relative to verification, $Ac_w/\kappa$},
  ylabel={Author reputation $\rho$},
  xtick={0,0.5,1,1.5,2},ytick={0,0.667,1},yticklabels={0,$\bar\rho$,1}]
\addplot[name path=L,black,thick,domain=0:0.6667,samples=80,variable=\c]
  ({(1-1.5*c)+0.5*((1-c/2)^2-c^2)},{c});
\addplot[name path=H,black,thick,domain=0:0.6667,samples=80,variable=\c]
  ({(1-1.5*c)+0.5*((1-c/2)^2-c^2)},{1-c/2});
\addplot[name path=L2,black,thick] coordinates {(1.5,0) (2,0)};
\addplot[name path=H2,black,thick] coordinates {(1.5,1) (2,1)};
\addplot[black!15] fill between[of=L and H];
\addplot[black!15] fill between[of=L2 and H2];
\draw[black!50,dashed] (axis cs:1.5,0) -- (axis cs:1.5,1);
\node[font=\small] at (axis cs:0.75,0.75) {verify (referee)};
\node[font=\small] at (axis cs:0.35,0.92) {trust (accept)};
\node[font=\small] at (axis cs:0.35,0.25) {ignore (desk reject)};
\node[font=\small,align=center] at (axis cs:1.75,0.5) {attention\\slack:\\verify all};
\node[font=\footnotesize,anchor=west] at (axis cs:1.15,0.12) {$\rho_L$};
\node[font=\footnotesize,anchor=west] at (axis cs:1.15,0.93) {$\rho_H$};
\end{axis}
\end{tikzpicture}
\caption{Who is read, or refereed. As writing becomes cheaper (to the right), the band of verified reputations shrinks from both sides and collapses to $\bar\rho$ in the costless writing limit, as Propositions \ref{Prop Reading} and \ref{Prop Equilibrium} show. Uniform $f$, $V=1$, $D=2$, $b_0=b=1$.}
\label{Fig Bands}
\end{figure}

\begin{figure}[h!]
\centering
\begin{tikzpicture}
\begin{axis}[width=0.8\textwidth,height=0.55\textwidth,
  xmin=0,xmax=2,ymin=0,ymax=1,axis lines=left,
  xlabel={Technological progress $t$},
  ylabel={Author reputation $\rho$},
  xtick={0,0.693,1,1.5,2},xticklabels={0,$\ln 2$,1,1.5,2},
  ytick={0,0.667,1},yticklabels={0,$\bar\rho$,1}]
\addplot[name path=PL,black,thick,domain=0:0.35179,samples=60,variable=\c]
  ({ln(((1-1.5*c)+0.5*((1-c/2)^2-c^2))/0.75)},{c});
\addplot[name path=PH,black,thick,domain=0:0.35179,samples=60,variable=\c]
  ({ln(((1-1.5*c)+0.5*((1-c/2)^2-c^2))/0.75)},{1-c/2});
\addplot[name path=PL2,black,thick] coordinates {(0.6931,0) (2,0)};
\addplot[name path=PH2,black,thick] coordinates {(0.6931,1) (2,1)};
\addplot[black!15] fill between[of=PL and PH];
\addplot[black!15] fill between[of=PL2 and PH2];
\addplot[black,thick,dashed,domain=0.35179:0.62581,samples=60,variable=\c]
  ({ln(0.75/((1-1.5*c)+0.5*((1-c/2)^2-c^2)))},{c});
\addplot[black,thick,dashed,domain=0.35179:0.62581,samples=60,variable=\c]
  ({ln(0.75/((1-1.5*c)+0.5*((1-c/2)^2-c^2)))},{1-c/2});
\node[font=\small,align=center] at (axis cs:1.45,0.3) {verification AI:\\every paper is verified};
\node[font=\small] at (axis cs:0.45,0.62) {writing AI};
\end{axis}
\end{tikzpicture}
\caption{The race between writing and reading of Proposition \ref{Prop Race}. The cost of writing falls as $c_w=e^{-t}$ in both scenarios. If AI cuts the attention that a verification requires twice as fast, $\kappa=e^{-2t}$, the verified band widens until every paper is verified at $t=\ln 2$ (solid, shaded). If it does not, $\kappa=1$, the band collapses to $\bar\rho$ (dashed). Parameters as in Figure \ref{Fig Bands}, with $A=0.75$.}
\label{Fig Race}
\end{figure}
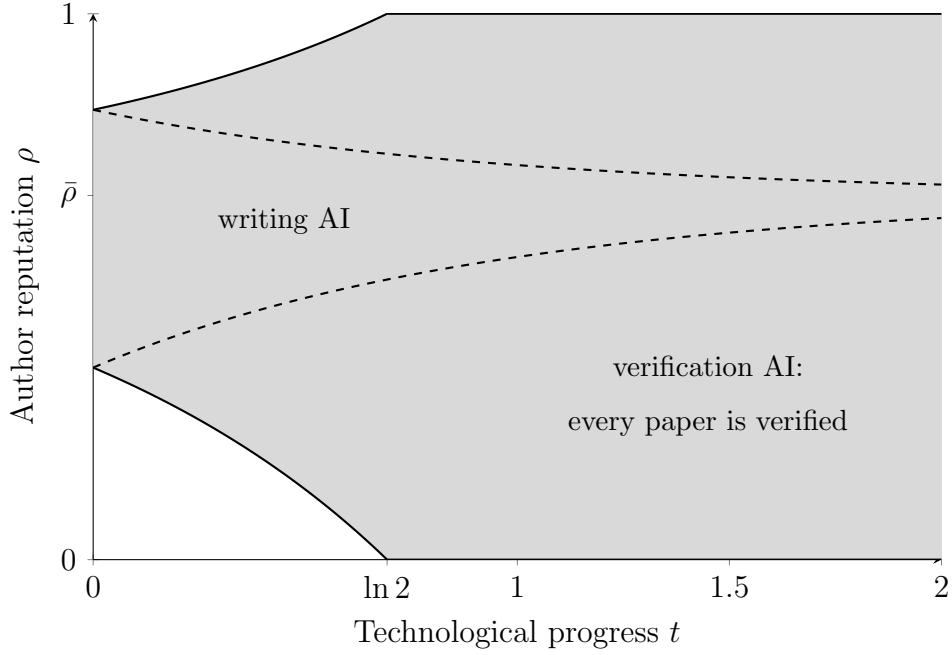

By Proposition \ref{Prop Race}, the reading rule depends only on $Ac_w/\kappa$, and Figure \ref{Fig Bands} plots it against this ratio, with the axis reversed so that cheaper writing moves the economy to the right. Left of the dashed line, attention is slack and every paper is verified. From there on, the lower boundary $\rho_L$ rises, so that outsiders are ignored, and the upper boundary $\rho_H$ falls, so that insiders are trusted. At the right edge, writing is free, and the band has collapsed to $\bar\rho$.

Figure \ref{Fig Race} adds time. In both scenarios, the cost of writing falls at the same rate; what differs is whether readers' capacity to check keeps pace. Both start at $Ac_w/\kappa=0.75$. With writing AI alone, the economy moves to the right in Figure \ref{Fig Bands}, and the band collapses toward $\bar\rho$. With AI that cuts $\kappa$ twice as fast, it moves to the left: outsiders are read again, insiders are checked again, and from $t=\ln 2$ on every paper is verified.

\FloatBarrier
\section{AI Summaries}\label{Section Summaries}

Scarce attention makes summaries and AI review attractive, and the cost of writing determines the composition of the literature. Suppose that readers, instead of reading papers, can ask an AI to summarize the literature consensus on a question. A \textit{summary} requires the same attention however many papers it covers, and we normalize this attention to zero, so that the flood of papers no longer exhausts readers' attention.

Consider a question with a true answer $\theta\in\{0,1\}$ and a large number of papers on it. A fraction $1-s$ of these papers is written carefully: each reports the correct answer with probability $q\in(1/2,1)$, independently of the others. The remaining fraction $s$ is written with the same AI. These papers share the AI's errors: they all report the same answer, which is correct with probability $p\in(1/2,1)$.\footnote{\citet{KimGarg2025} provide empirical support for shared errors: evaluating more than 350 language models, they find that on one leaderboard two models agree $60\%$ of the time when both err, and that larger, more accurate models have highly correlated errors, even across providers.} We call these papers AI-written. The term refers to papers that adopt the AI's answer without independent work. To make production consistent with Section \ref{Section Model}, let a careful paper cost $c_w+c_I$, where $c_I>0$ is the cost of independent work, which AI does not lower, while an AI-written paper costs only $c_w$. With this cost function and a common use probability $u$, a mass $m\in(0,1)$ of authors who copy the AI writes $(b_0+bu)/c_w$ papers each, and the others write $(b_0+bu)/(c_w+c_I)$ each, by (\ref{Eq Papers}). The share of AI-written papers is then $s(c_w)=m(c_w+c_I)/(c_w+mc_I)$. Hence $s(c_w)$ is decreasing, with $s\to1$ as $c_w\to0$. We assume that the errors of careful papers are independent of each other and of the AI's error. The summary reports the majority answer of the literature, and a tie counts as an error. Majority rule is a benchmark in which the summary is faithful to the literature it reads. A summary that leans toward the AI's own answer is captured at a lower share than the threshold below, so the threshold is an upper bound for such summaries.

\begin{prop}[Captured Consensus]\label{Prop Capture}
Let $\bar s:=1-1/(2q)\in(0,1/2)$. In the limit of a large literature, the summary is correct with probability one if $s<\bar s$, and with probability $p$ if $s>\bar s$. Hence, if $s(c_w)<\bar s$ at high costs of writing, there is a cost $\bar c_w$ such that, for $c_w<\bar c_w$, the error rate of the summary jumps from $0$ to $1-p$. Its errors are then the same for all readers who use it.
\end{prop}

Below the threshold, careful papers outvote the AI's errors, and the summary is more reliable than any single paper. In a finite literature, the probability that they do so converges to one exponentially in the number of papers. The threshold is a tipping point, as in \citet{BergstromGross2025}, but it arises from correlated errors rather than from congestion. A captured summary is like an unverified paper by an author of reputation $p$: by Proposition \ref{Prop Reading}, readers use it if and only if $p\ge\bar\rho$.

\begin{rem}[Provenance]\label{Rem Provenance}
If the AI can identify AI-written papers and gives them the relative weight $w\in[0,1]$, the summary remains correct as long as $sw/(sw+1-s)<\bar s$. As $s\to1$, this requires $w\to0$: the AI must discard papers that share its own errors, which requires it to tell them apart. When writing is free, the provenance of a paper becomes the information a summary needs, and reputation re-enters through the AI's weights.
\end{rem}

\begin{rem}[Partial Dependence]\label{Rem Partial}
Suppose that each AI-written paper adopts the AI's answer only with probability $\gamma\in(0,1]$, and otherwise reports an independent answer that is correct with probability $q$. When the AI is wrong, the share of correct papers is then $q(1-\gamma s)$, so that the summary is correct with probability one if and only if $\gamma s<\bar s$.
\end{rem}

\subsection{AI Peer Review}\label{Section Review}

Suppose that journals, whose referees verify an ever smaller band of papers under the reading rule of Proposition \ref{Prop Reading}, eventually delegate review to the AI, and that every paper must pass AI review to enter the summary. An AI referee judges a paper against what it has learned: it accepts papers that agree with its own answer, and accepts papers that disagree with it only with probability $r\in[0,1]$, the probability that the evidence in a dissenting paper persuades it. Since $r$ does not depend on whether a dissenting paper is correct, the AI screens on agreement. Dissenting papers are the original research of the model: when the AI is wrong, they are the papers that are right.

\begin{prop}[Filtered Originality]\label{Prop Review}
Let $\bar s(r):=1-1/(q(1+r))$. With AI review, the summary is correct with probability one if $s<\bar s(r)$, and with probability $p$ if $s>\bar s(r)$. The threshold $\bar s(r)$ is increasing in $r$, equals $\bar s$ for $r=1$, and is negative for $r<(1-q)/q$. In the latter case, the summary repeats the AI even if no paper is written with AI.
\end{prop}

\section{Policy Implications: Preprint Servers}\label{Section Policy}

\begin{figure}[h!]
\centering
\begin{tikzpicture}
\begin{axis}[width=0.75\textwidth,height=0.45\textwidth,
  xmin=0,xmax=1,ymin=0,ymax=0.5,axis lines=left,
  xlabel={Probability $r$ that the AI referee accepts dissent},
  ylabel={Share $s$ of AI-written papers},
  ytick={0,0.167,0.375,0.5},yticklabels={0,0.167,0.375,0.5},
  xtick={0,0.25,0.5,1}]
\addplot[name path=S,black,thick,domain=0.25:1,samples=60] {1-1/(0.8*(1+x))};
\addplot[name path=Top,draw=none,domain=0:1] {0.5};
\addplot[black!15] fill between[of=Top and S,soft clip={domain=0.25:1}];
\fill[black!15] (axis cs:0,0) rectangle (axis cs:0.25,0.5);
\draw[black!60,dotted,thick] (axis cs:0,0.375) -- (axis cs:1,0.375);
\addplot[only marks,mark=*,mark size=1.8pt] coordinates {(0.5,0.1667) (1,0.375)};
\draw[-{Stealth},thick] (axis cs:0.5,0.3) -- (axis cs:0.98,0.3);
\node[font=\footnotesize,anchor=south] at (axis cs:0.66,0.303) {preprint server};
\node[font=\small] at (axis cs:0.5,0.46) {consensus repeats the AI};
\node[font=\small] at (axis cs:0.78,0.08) {consensus correct};
\node[font=\footnotesize,anchor=south west] at (axis cs:0.02,0.378) {no acceptance rule helps above $\bar s$};
\end{axis}
\end{tikzpicture}
\caption{Captured consensus, from Propositions \ref{Prop Capture} and \ref{Prop Review}. Below the curve $\bar s(r)=1-1/(q(1+r))$, careful papers outvote the AI; above it, the summary repeats the AI. A lower $r$ reduces the threshold, and for $r<(1-q)/q$, review alone captures the consensus. A summary that reads unrefereed preprints moves from $r$ to $1$ (arrow) and remains correct up to $\bar s=1-1/(2q)$ (dotted), above which no acceptance rule helps ($q=0.8$).}
\label{Fig Capture}
\end{figure}
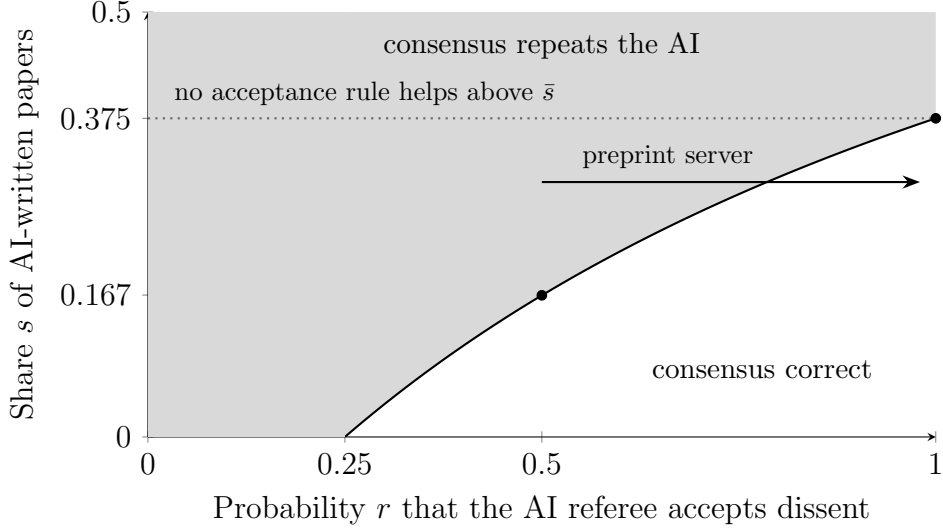

The simplest remedy for the filter of AI review is an unrefereed preprint server from which the summary draws. A summary that reads all preprints accepts dissent with probability $r=1$, so that, by Proposition \ref{Prop Review}, it is correct whenever $s<\bar s=1-1/(2q)$, the threshold without review. No acceptance rule can do better, since $\bar s(r)$ is increasing in $r$, as Figure \ref{Fig Capture} illustrates. When the AI is right, incorrect dissent on the server does not change the majority and thus does not harm the summary. Review and summary then serve distinct purposes: review certifies individual papers for human readers, the summary aggregates the unfiltered record, and the preprint label indicates which papers have not been checked.

The unfiltered channel is, however, itself exposed to the flood. In October 2025, arXiv stopped accepting review articles and position papers in computer science that had not passed peer review, following a surge of AI-generated submissions.\footnote{arXiv, ``Attention Authors: Updated Practice for Review Articles and Position Papers in arXiv CS Category,'' arXiv blog, October 31, 2025.} In our model, such a requirement imports the very filter of the review process it relies on.

\FloatBarrier
\section{Discussion}\label{Section Discussion}

The model identifies two consequences of cheaper research writing. First, as writing becomes cheaper, the number of papers rises while readers' attention remains fixed. Verification is rationed to an increasingly narrow band of intermediate-reputation authors: outsiders are ignored, and established authors are trusted without being checked. In the costless writing limit, the share of verified papers vanishes, and whether a paper is used depends on its author's reputation alone.

Second, AI summaries allow readers to bypass the attention bottleneck, but their reliability depends on the independence of the underlying research. In a large literature, a majority summary recovers the truth when sufficiently many papers reach independent, better-than-random judgments. Once enough papers adopt the same AI's answer, their common errors capture the majority, and the summary repeats the AI's mistakes. AI review that favors agreement compounds this problem: when the AI is wrong, the dissenting papers it filters out are precisely those that could correct the consensus. A summary that draws on unrefereed preprints removes this filter but cannot always restore a reliable consensus.

The share of authors who adopt the AI's views is exogenous in our model. If it were endogenous, an AI that aggregated research perfectly would remove the incentive to produce research, as in \citet{GrossmanStiglitz1980}. Independent work would thus need a reward that adopting the AI's views does not earn.

Finally, the mechanism we describe can run in reverse: If AI lowers the cost of verifying papers faster than the cost of writing, the verified band widens, such that outsiders are read and established authors are checked again. Likewise, a summary that draws on unrefereed preprints and identifies AI-written papers remains reliable as long as enough original papers remain to outvote the AI's errors.

\hspace{4cm}

\noindent\textbf{Declarations}\\
\textit{Funding:} This research received no external funding. \textit{Competing interests:} The author has no competing interests. \textit{Data availability:} Not applicable; no data were used. \textit{Use of AI:} AI was used to check language and grammar.

\newpage
\appendix
\section{Proofs}\label{Appendix Proofs}

\begin{proof}[Proof of Proposition \ref{Prop Reading}]
Verifying is at least as good as ignoring if and only if $\rho\ge c_v/V$, and at least as good as trusting if and only if $\rho\le1-c_v/D$. Trusting is at least as good as ignoring if and only if $\rho\ge\bar\rho$. If $c_v<\bar c$, then $c_v/V<\bar\rho<1-c_v/D$, which gives the three bands. If $c_v\ge\bar c$, then $c_v/V\ge1-c_v/D$, so that verification is never strictly optimal, and readers compare ignoring with trusting.
\end{proof}

\begin{proof}[Proof of Proposition \ref{Prop Equilibrium}]
$\Phi$ is continuous, and for $c_v\in(0,\bar c)$,
$$\Phi'(c_v)=-\frac1V\Big(b_0+b\frac{c_v}{V}\Big)f\Big(\frac{c_v}{V}\Big)-\frac1D\Big(b_0+b\Big(1-\frac{c_v}{D}\Big)\Big)f\Big(1-\frac{c_v}{D}\Big)<0,$$
with $\Phi(\bar c)=0$, since the limits of integration coincide at $\bar c$. By (\ref{Eq Papers}), verified papers use $\kappa\Phi(c_v)/c_w$ units of attention. If $\kappa\Phi(0)/c_w\le A$, attention is slack at $c_v=0$, and $\lambda=0$. Otherwise, $\lambda>0$, attention binds, and (\ref{Eq Equilibrium}) has a unique solution $c_v\in(0,\bar c)$, which is decreasing in $Ac_w/\kappa$ and thus in $c_w$. As $c_w\to0$, the right-hand side of (\ref{Eq Equilibrium}) converges to $0$, so that $c_v\to\bar c$, and $\rho_L,\rho_H\to\bar\rho$. Since $b_0\le b_0+bu\le b_0+b$, the number of papers $N=\int_0^1n(\rho)f(\rho)\,d\rho$ satisfies $b_0/c_w\le N\le(b_0+b)/c_w$, while the number of verified papers equals $A/\kappa$. Hence the share of verified papers is at most $Ac_w/(\kappa b_0)\to0$.
\end{proof}

\begin{proof}[Proof of Proposition \ref{Prop Race}]
Equation (\ref{Eq Equilibrium}) and the slackness condition involve $(c_w,\kappa,A)$ only through $Ac_w/\kappa$, and the cutoffs depend on $c_v$ alone. If attention binds, $c_v$ is strictly decreasing in $Ac_w/\kappa$, so that $\rho_L=c_v/V$ falls and $\rho_H=1-c_v/D$ rises if and only if $c_w/\kappa$ rises, i.e.\ if and only if $\kappa$ falls proportionally faster than $c_w$.
\end{proof}

\begin{proof}[Proof of Proposition \ref{Prop Capture}]
By the law of large numbers, the share of careful papers that report the correct answer converges to $q$. If the AI's answer is correct, the share of papers reporting the correct answer is $(1-s)q+s>1/2$. If the AI's answer is wrong, this share is $(1-s)q$, which exceeds $1/2$ if and only if $s<\bar s$. Hence the summary is always correct if $s<\bar s$, and correct exactly when the AI is, with probability $p$, if $s>\bar s$. Since $s(c_w)$ is continuous and decreasing with $s\to1$ as $c_w\to0$, there is a $\bar c_w$ with $s(\bar c_w)=\bar s$. All readers who use the summary observe the same answer, and thus the same error.
\end{proof}

\begin{proof}[Proof of Proposition \ref{Prop Review}]
Per unit of papers, AI-written papers and careful papers that agree with the AI pass review, and careful papers that disagree pass with probability $r$. If the AI is correct, the accepted share reporting the correct answer is $(s+(1-s)q)/(s+(1-s)q+(1-s)(1-q)r)>1/2$, since $q>1/2\ge(1-q)r$. If the AI is wrong, correct papers are careful papers that disagree with it, and their accepted share is $(1-s)qr/\big((1-s)qr+s+(1-s)(1-q)\big)$, which exceeds $1/2$ if and only if $(1-s)(q(1+r)-1)>s$, i.e.\ if and only if $s<\bar s(r)$. The threshold $\bar s(r)=1-1/(q(1+r))$ is increasing in $r$, equals $1-1/(2q)=\bar s$ at $r=1$, and is negative if and only if $q(1+r)<1$, i.e.\ $r<(1-q)/q$; then $s<\bar s(r)$ fails for every $s\ge0$.
\end{proof}

\newpage
\addcontentsline{toc}{section}{References}
\bibliographystyle{apalike}
\bibliography{References}

\end{document}